# A Novel Sub-Nyquist Multiband Signal Detection Algorithm for Cognitive Radio

Kai Cao, Peizhong Lu, Yan Zou, and Lin Ling

*Abstract*: Wideband spectrum sensing (WSS) is an essential technology for cognitive radio. However, the sampling rate is still a bottleneck of WSS. Several sub-Nyquist sensing methods have been proposed. These technologies deteriorate in the low signal to noise ratio (SNR) regime or suffer high computational complexity. In this paper, we propose a novel sub-Nyquist WSS method based on Multi-coset (MC) sampling. We design a simple SNR-robust and low-complexity multiband signal detection algorithm. In particular, the proposed method differs the commonly used detection algorithms which are based on energy detection (ED), matched filter (MF) or cyclostationary detection (CD). We exploit the linear recurrent relation between the locations of nonzero frequencies and the DFT of the arithmetic-shifted subsampled signals. These relations can be uniquely expressed by a series of the so-called frequency locator polynomials (FLPs). The scalar of the relations is related to the bandwidths of the subsignals. Basing on this, we propose a detector for sparse multiband signals along with the method estimating carrier frequency and bandwidth. The detector does not require priori knowledge about the frequency locations of the signals of interest. Moreover, it has lower complexity of both samples and computation compared to CD in sparse case. Experimental results show the detector outperforms ED in the sub-Nyquist regime especially in low SNRs.

*Keywords—Spectrum sensing, sub-Nyquist sampling, locator polynomial, cognitive radio*

## 1 Introduction

Cognitive Radio (CR) is proposed for potentially solving the problem of spectrum overcrowdedness and bridging the scarcity of spectral resources and their sparse nature [1]. Spectrum sensing [2] is the key mission of CR, which has strict requirements on the performance of both software and hardware.

Matched filter (MF) [3], [4] is the optimal linear filter in the sense of maximizing the signal to noise ratio (SNR) in the presence of additive stochastic noise. It requires full knowledge of the signals. In contrast, energy detection (ED) [5] is the simplest approach and does not require any priori knowledge. However, ED deteriorates at low SNR. Cyclostationary detection (CD) [6] [7] is proposed as a compromise between both methods. This approach is more robust to noise compared to ED but assumes the signal of our interest exhibits cyclostationarity.

High Nyquist rate of wideband signal is a heavy burden even for the state-of-art analog-to-digital converters (ADCs). Moreover, the big samples data is also a challenge for the real-time signal processing unit. Consider that the spectrum is not always fully occupied, namely sparse. To efficiently sample the sparse wideband signals, several sub-Nyquist sampling methods have been proposed [8]-[11]. A certain aliasing filter is generally used such that the sub-signals aliases in the baseband. Then, through sampling by several parallel sampling channels equipped with low-rate ADCs and processing with aliasing resolving algorithm, the original signal can be recovered. Among these methods, multi-coset (MC) [12]-[14] sampling and Modulated Wideband Converter (MWC) [15] are popularly used [17]. MC subsamples the signal with distinct intervals and obtains subsampled signals with different phases. In comparision, MWC modulates both phase and amplitude of the signals. The total sampling rates of both methods are equal to $f_{lan}$ and $\min\{2f_{lan}, f_{nyq}\}$ respectively when the frequency locations are known and unknown (blind case), where $f_{lan}$ is the Landau rate and $f_{nyq}$ is the Nyquist rate [10].

In this paper, we focus on blind detection of the sparse multiband signals in the sub-Nyquist regime. The candidate methods solving this problem in the published researches include ED and CD. CD is more favorable for its excellent performance in the low SNR regimes. Zhi Tian *et. al.* [18] [19] propose a cyclostationary detection based algorithm for wideband sensing using the latest achievement in compressed sensing. The method is reported to be robust to low SNR conditions and unpredictable noise uncertainty in wireless networks. Cohen and Eldar propose a multiband signal detecting algorithm based on sub-Nyquist cyclostationary detection[19]. This method is composed of three steps (or algorithms): MWC (for Sub-Nyquist Sampling), CTF (for Support Recovery) and OMP (orthogonal matching pursuit, for Cyclic Spectrum Recovery). The three sub-technologies have been proved to perform well in their previous researches. In this work, the authors derive the lower bound of sampling rates for the cyclic spectrum recovery in the presence of noise. The stated bounds are respectively $8/5f_{lan}$ when the spectrum is sparse and $4/5f_{nyq}$ without any sparsity constraints. The better performance of cyclostationary detection compared to ED in the sub-Nyquist regime is also pronounced in the reference.

Study [21] presents a sparse spectrum sensing and decoding algorithm named AD-BigBand, which is based on equi-space shifted sub-Nyquist sampling. The frequency locator polynomial (FLP) is defined and used to rapidly recovery the

---

Kai Cao is with the National Digital Switching Engineering Technological Research Center, Zhengzhou 450002, P.R. China. e-mail: ck1988@mail.ustc.edu.cn

Peizhong Lu is with the Department of Computer Science and Engineering, Fudan University, Shanghai, 200433, P.R. China. e-mail: (pzlu@fudan.edu.cn)

Yan Zou and Lin Ling are with the Southwest Inst. of Electron & Telecom. Tech., Shanghai 200434, P.R. China e-mail: (yanzouzou@163.com, lynn-00@163.com)

wideband signal. FLP is similar to the error locator polynomial in decoding theory [22]. Our work is based on [21], however we have the following improvements.
   a. We provide the algebraic and statistical methods for comprehensively analyzing the FLP. The evaluation of FLP is used as the detection measurement and its probability density function is derived.
   b. We show the detection probability increase with the growing $d$, which is the scale of the linear relations determined by an identical FLP, from theoretical and experimental aspects.
   c. We demonstrate that the wideband signal can be detected even in very low SNRs with low false alert probability.

In this paper, we propose a simple and practical algorithm of multiband signal blind detection for wideband spectrum sensing. Our method is different from energy detection and cyclostationary detection in that the proposed algorithm does not need to recovery the power or cyclic spectrum. We subsample the original signal with MC sampling technology and the time-shifts of the sampling channels are equispaced. We provide an algorithm based on FLP to locate the nonzero frequencies in the original spectrum thus to completing detecting signals and estimating their carrier frequencies and bandwidths. In the presence of noise, we estimate the FLPs by using least square (LS) method and then design a detector based on the polynomial evaluations of the candidate roots (corresponding to the frequency locations). Experimental results show that the proposed detector has high detection probability even in the low SNRs.

The main contributions of this paper are as follows:
   a. Lower sampling rate. Our method requires lower sampling rate in the sparse case compared to CD [19]. The total sampling rate in our scheme is $\frac{N_S+1}{N_S} f_{lan}$ which is lower than the bound derived in [19] when $N_S > 1$, where $N_S$ is the number of the sub-bands. This result also indicates that our algorithm has lower sample complexity.
   b. Lower computational complexity. We show that the computing cost of the proposed algorithm is $O(K)$, which is lower than the CD based methods, where $\kappa$ is the spectral sparsity.
   c. Looser requirement for priori knowledge. MF requires the full knowledge of the target signal and CD assumes the signal of interest exhibits cyclostationarity. In contrast, our algorithm has only one assumption that the minimal bandwidth of the sub-signals is larger than $N_S$. This is a reasonable assumption since the communication signals generally has bandwidths up the order of kHz.
   d. Robustness to noise. The experimental results show that our method can robustly detect the signals even in low SNR regime where ED performs poorly.

This paper is organized as follows. In Section II, we review the multiband signal model and the multi-coset sampling. Section III presents the proposed algorithm as well as the related theoretical derivation and proof. Comparison between the detection algorithms is given in Section IV. Numerical experiments and analysis are given in Section V. Finally, Section VI concludes the paper.

## 2 Multiband Signal Model and Sampling Method

### 2.1 Multiband Signal Model and Problem Statement

In the theoretical part of this article, the signal is assumed to be complex. Suppose the wideband signal is supported on $\mathcal{F} = [0, f_{\max}]$ and composed of up to $N_S$ subsignals which are distinct in the frequency domain. Let the $i-th$ subsignal be

$$s_i(t) = \sum_{n \in Z} d_i[n] g_i(t - nT_i) e^{j2\pi f_i^c t},$$

where $\{d_i[n]\}$ is the symbol sequence, $g_i(t)$ is the pulse shaping function and $f_i^c$ is the carrier frequency of $s_i(t)$. $s_i(t)$ is band limited to $[l_i, u_i)$. Thus the bandwidth of $s_i(t)$ is $u_i - l_i$ and $f_i^c = \frac{u_i + l_i}{2}$. Then we obtain $f_{nyq} = f_{\max}$ and $f_{lan} = \sum_{i=0}^{n_S-1}(u_i - l_i)$. Here, we assume $l_0 \leq u_0 \leq \cdots \leq l_{n_{sig}-1} \leq u_{n_s-1}$. The corrupted signal can be expressed as

$$x(t) = \sum_{i \in [n_S]} s_i(t) + n(t),$$

where $n(t)$ is the additive Gaussian white noise.

The problem of wideband spectrum sensing is equivalent to estimate $N_S$, the number of the subsignals and their carrier frequencies and bandwidths. Let the bandwidth of each subsignal not exceed $B_{\max} = \max_i(u_i - l_i)$ and the number of them not exceed $N_S$. Consider the case that the spectrum is sparse, namely, $B_{\max} N_S \ll f_{\max} = f_{nyq}$. Generally, $f_{\max}$ is more than GHz in WSS. The Nyquist sampling is not advisable. The popular sampling method is MC and MWC. In this paper, MC is used.

### 2.2 Multi-coset Sampling

In practice, MC is realized by using several parallel ADCs at same rate and with distinct intervals to uniformly sample the signal of interest. Let the sampling rate of each ADC be $f_{nyq}/\alpha$, where $\alpha$ is the subsampling ratio. Assume the number of ADCs (or cosets) is $r, r \leq \alpha$ and the interval of the $i-th$ coset is $\tau_i T_{nyq}$, where $i = 0, ..., r-1, T_{nyq} = 1/N$, $0 \leq \tau_0 < \tau_1 < \cdots < \tau_{r-1} < \alpha - 1$ and $N = f_{nyq}$ is the FFT point. The discrete sample sequence can be written as

$$y_i(n) = x(n\alpha + \tau_i), n = 0, ..., N/\alpha - 1, i = 0, ..., r-1. \tag{1}$$

The discrete Fourier transform (DFT) of $y_i(n)$ is

$$Y_i(m) = \sum_{n=0}^{N/\alpha-1} y_i(n) e^{-j2\pi \frac{mn}{N/\alpha}} = \sum_{n=0}^{N/\alpha-1} x(\alpha n + \tau_i) e^{-j2\pi \frac{m}{N}(\alpha n + \tau_i - \tau_i)} = e^{-j2\pi \frac{m}{N}\tau_i} \sum_{n=0}^{N/\alpha-1} x(\alpha n + \tau_i) e^{-j2\pi \frac{m}{N}(\alpha n + \tau_i)} . \quad (2)$$

For

$$\sum_{l=0}^{\alpha-1} e^{-j2\pi l \frac{n-\tau_i}{\alpha}} = \begin{cases} \alpha, & n = k\alpha + \tau_i, k = 0, \pm 1, \cdots \\ 0, & \text{otherwise} \end{cases},$$

then (2) can be written as

$$Y_i(m) = e^{-j2\pi \frac{m}{N}\tau_i} \sum_{n=0}^{N-1} \left( x(n) e^{-j2\pi m \frac{n}{N}} \sum_{l \in [\alpha]} \frac{1}{\alpha} e^{-j2\pi l \frac{n-\tau_i}{\alpha}} \right)$$

$$= \sum_{l=0}^{\alpha-1} \frac{1}{\alpha} e^{j2\pi \left(m + l\frac{N}{\alpha}\right)\tau_i} \underbrace{\sum_{n \in [N]} x(n) e^{-j2\pi \frac{n}{N}(m + l\frac{N}{\alpha})}}_{X\left(m + l\frac{N}{\alpha}\right)} = \frac{1}{\alpha} \sum_{l=0}^{\alpha-1} X\left(m + l\frac{N}{\alpha}\right) e^{j2\pi \left(m + l\frac{N}{\alpha}\right)\frac{\tau_i}{N}} . \quad (3)$$

Let $m = f, f \in [0, N/\alpha), f_l = f + l\frac{N}{\alpha}$. Rewrite (3) as

$$Y_i(f) = \sum_{n=0}^{N/\alpha-1} y_i(n) e^{-j2\pi \frac{mn}{N/\alpha}} = \frac{1}{\alpha} \sum_{l=0}^{\alpha-1} X(f_l) e^{j2\pi f_l \frac{\tau_i}{N}} . \quad (4)$$

Relation (4) indicates that the subsampled spectrum is an aliasing version of the original spectrum with a phase rotation which is related to the time shift. If we regard the aliasing process as frequency hashing, then each frequency in the aliased spectrum corresponds to a hashing bucket. The $i$-th bucket is $\mathbf{H}_i = \{f \in \mathbb{Z} \mid 0 \le f < N, f = i \bmod N/\alpha\}$, where $i = 0, \ldots, N/\alpha - 1$.

We consider the following case that the spectrum is sparse. Let $f_{i,j} = i + (l_{i,j} - 1)N/\alpha$ denotes the $j-th$ nonzero frequencies hashed into $\mathbf{H}_i$, where $i \in [0, N/\alpha), j \in [0, n_S)$ and $l_{i,j} \in \mathbf{I}_i = \left\{ h \mid X\left(i + (h-1)\frac{N}{\alpha}\right) \ne 0 \right\}, i \in [0, N/\alpha)$. Obviously, the elements in $\mathbf{I}_i$ and the nonzero frequencies in $\mathbf{H}_i$ are one-to-one mapping. It means the carrier frequencies and bandwidths of all the transmissions in the wideband can be determined if $\mathbf{I}_i$ is obtained. Then, the signal detection problem is converted into determining $\mathbf{I}_i, i \in [0, N/\alpha)$, namely, the location sets of the nonzero frequencies.

Let $n_i = |\mathbf{I}_i| \le N_S < r, i = 0, \ldots N/\alpha - 1$, then we can write (4) as the matrix form

$$\underbrace{\begin{bmatrix} Y_0(i) \\ Y_1(i) \\ \vdots \\ Y_{r-1}(i) \end{bmatrix}}_{\mathbf{Y}_i} = \frac{1}{\alpha} \underbrace{\begin{bmatrix} 1 & \cdots & 1 \\ w^{f_{i,0}\tau_1} & \cdots & w^{f_{i,N_S-1}\tau_1} \\ \vdots & \ddots & \vdots \\ w^{f_{i,0}\tau_{r-1}} & \cdots & w^{f_{i,N_S-1}\tau_{r-1}} \end{bmatrix}_{r \times n_S}}_{\mathbf{W}_i} \underbrace{\begin{bmatrix} X_{f_{i,0}} \\ X_{f_{i,1}} \\ \vdots \\ X_{f_{i,N_S-1}} \end{bmatrix}_{N_S \times 1}}_{\mathbf{X}_i}, \quad (5)$$

where $w \triangleq e^{j2\pi/N}$ and $|\cdot|$ denotes the $l_0$ norm of a set. The above relations can be seen from fig. 1. In the following section, we show that if the time shifts are equispaced, there exists linear recurrent relations between the DFT of the subsampled signals and the locations of the occupied bands in the original spectrum.

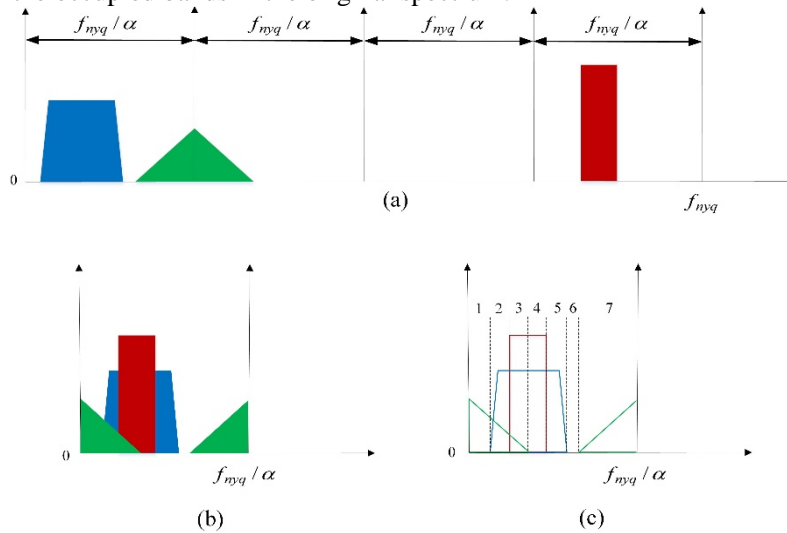

Fig. 1 Original and subsampled spectrum. The number of the total nonzero frequencies hashed into the buckets in the seven districts of figure

(c) are respectively $n_r = 1, 2, 3, 2, 1, 0, 1$.

## 3 Multiband Signal Detection Algorithm

In this section, we give the detection algorithm for sparse multiband signal based on multi-coset sampling. This method constructs a series of FLPs to characterize the linear recurrent relations between the subsampled spectrums and occupied frequencies, thus to sense the wideband spectrum. The FLPs are estimated from sufficient linear equations.

### 3.1 Frequency Locator Polynomial

For the sparse spectrum subsampled by $\alpha \times$, only a little bit of the aliased frequencies are of nonzero spectrum value. For the bucket $\mathbf{H}_i$ and the corresponding nonzero frequency location set $\mathbf{I}_i$, the frequency locator polynomial is defined as[21]:

$$G_i(z) = \prod_{l_{i,j} \in \mathbf{I}_i} (1 - w^{-f_{i,j}\tau} z), i = 0, ... N/\alpha - 1,$$

where $f_{i,j} = i + (l_{i,j} - 1)N/\alpha$ and $\tau$ is a constant.

Then we have the following theorem.

**Theorem 1[21]:** Let $\tau_s = s\tau, \tau = c, s = 0,1,...,r-1$, where $c(r-1) < \alpha - 1$ and $c$ is a positive integer constant. Assume $n_i$, the number of nonzero frequencies in bucket $\mathbf{H}_i$ satisfies $n_i = |\mathbf{I}_i| \le N_S < r$, where $i = 0,..., N/\alpha - 1$. Let $G_i(z)$ be the FLP corresponding to $\mathbf{H}_i$, such that

$$G_i(z) = \sum_{s=0}^{n_i} a_{is} z^s ,\tag{6}$$

Then we have the following relations

$$\sum_{s=0}^{n_i} a_{is} Y_{s+t}(i) = 0, t = 0,..., r-1-n_i ,\tag{7}$$

where $a_{is} \in \mathbb{C}, a_{i0} = 1, i = 0,..., N/\alpha - 1$.

Since some of the relations in the proof of Theorem 1 will be used in the following sections, we borrow the proof from [21].

*Proof:* Let

$$G_i(z) = \prod_{l_{i,j} \in \mathbf{I}_i} (1 - w^{-f_{i,j}\tau} z) = \sum_{s=0}^{n_i} a_{is} z^s .\tag{8}$$

Then $\omega_{i,j} = w^{f_{i,j}\tau}$ are the roots of polynomial $G_i(z)$, and relation (5) can be written as

$$\underbrace{\begin{bmatrix} Y_0(i) \\ Y_1(i) \\ \vdots \\ Y_{r-1}(i) \end{bmatrix}}_{\mathbf{Y}_i} = \frac{1}{\alpha} \underbrace{\begin{bmatrix} 1 & \cdots & 1 \\ \omega_{i,0}^1 & \cdots & \omega_{i,n_i-1}^1 \\ \vdots & \ddots & \vdots \\ \omega_{i,0}^{r-1} & \cdots & \omega_{i,n_i-1}^{r-1} \end{bmatrix}}_{\mathbf{W}_i} \underbrace{\begin{bmatrix} X_{f_{i,0}} \\ X_{f_{i,1}} \\ \vdots \\ X_{f_{i,n_i-1}} \end{bmatrix}}_{\mathbf{X}_i}.\tag{9}$$

Let $\vec{a}_i^{(t)} = [\overbrace{0,...,0}^{t}, a_{i0}, a_{i1},..., a_{in_i}, \overbrace{0,...,0}^{r-n_i-t}], t \in [r-1-n_i]$, then

$$\vec{a}_i^{(t)} \mathbf{W}_i = \frac{1}{\alpha} \left[ \omega_{i,0}^t G_i(\omega_{i,0}), \omega_{i,1}^t G_i(\omega_{i,1}),..., \omega_{i,n_i-1}^t G_i(\omega_{i,n_i-1}) \right] = 0 .$$

Thus

$$\sum_{s=0}^{n_i} a_{is} Y_{s+t}(i) = \vec{a}_i^{(t)} \mathbf{Y}_i = \vec{a}_i^{(t)} \mathbf{W}_i \mathbf{X}_i = 0, t = 0,..., r-1-n_i ,$$

for $i = 0,..., N/\alpha - 1$. The theorem is proved.

In the following part of this paper, we set $\tau_s = s, s = 0,1,...,r-1$.

Theorem 1 indicates that the nonzero frequencies set $\mathbf{I}_i$ can be uniquely determined by a FLP. In reverse, we can use the linear relations to obtain the FLPs and $\mathbf{I}_i$.

### 3.2 Approximating the FLPs

The approximation of the FLPs is equivalently to estimate their coefficients $\mathbf{a}_i = [a_{i1}, a_{i2},..., a_{in_i}], i = 0,..., N/\alpha - 1$. Note that $\mathbf{a}_i$ can be obtained by solving equations (7). If $r \ge 2n_i$, (7) uniquely determines $\mathbf{a}_i$. If $r < 2n_i$, (7) is underdetermined and we need to construct another equations to approximate $\mathbf{a}_i$.

Assume $B_{\max}$, the maximal bandwidth of the sub-band signals satisfies $B_{\max} \le N/\alpha$, which can guarantee $n_i = |\mathbf{I}_i| \le N_S$ for $i = 0, ..., N/\alpha - 1$. Note the following facts for communication signals.

  a. A communication signal usually successively occupies several frequencies and the dimension is its bandwidth. Namely, in a sensing window, the spectrum of a signal is successive.
  b. The nonzero frequencies hashed into the adjacent buckets, e.g. $\mathbf{H}_i$ and $\mathbf{H}_{i+1}$, usually neighbor each other in the original frequencies. This can be seen from fig. 1c: for each hashing bins in district 3, the nonzero frequencies in it are all respectively contributed by the spectrum of the three signals (marked in three different colors) and they are correspondingly neighboring.

Based on the above facts, we have the following property.

**Property 1**: If we divide the subsampled spectrum into distinct districts based on the difference between the numbers of the nonzero frequencies hashed in each bucket. Namely, $\left[0, \frac{N}{\alpha}\right) = \bigcup_{s=1}^{n_D} \mathbf{d}_s = \bigcup_{s=1}^{n_D} [l_s, u_s)$, where $n_D$ is the total number of the districts (See fig. 1c, where the aliased spectrum is divided into seven districts). Then for the buckets in the same district, the corresponding sets of nonzero frequency locations are identical, namely $\mathbf{I}_{l_s} = \mathbf{I}_{l_s+1} = \cdots = \mathbf{I}_{u_s-1}$, $s = 1, 2, ..., n_D$.

By Property 1, for the adjacent buckets $\mathbf{I}_i$ and $\mathbf{I}_{i+1}$ in a specific district we have $\mathbf{I}_i = \mathbf{I}_{i+1}$. Then by (7) and [20], we have

$$\sum_{s=0}^{n_i} a_{is} Y_{s+t}(i) = 0, \sum_{s=0}^{n_i} a_{is} \theta^{-(s+t)} Y_{s+t}(i+1) = 0, t = 0, ..., r-1-n_i,$$

where $\theta = e^{j2\pi/N}$. Accordingly, for the district $\mathbf{d}_k$, we have

$$\sum_{s=0}^{n_i} a_{is} \theta^{-(s+t)m} Y_{s+t}(l_k + m) = 0, t = 0, ..., r-1-n_i, m = 0, ..., u_k - l_k - 1. \tag{10}$$

Relations (10) indicates that through multiplying specific coefficients to the DFT of the subsampled signals, the buckets in the same district determine an identical FLP. Accordingly, the frequency locations sets $\mathbf{I}_i, i = 0, ..., N/\alpha - 1$ can be uniquely determined by $n_D$ different FLPs.

Note in the practical processing, $\mathbf{d}_s, s = 1, ..., n_D$ is unknown, thus the relation scale of (10) is also unknown. Assume that $d$ successive buckets $\mathbf{H}_i, ..., \mathbf{H}_{i+d-1}$ corresponds an identical FLP, namely $\mathbf{a}_i = \cdots = \mathbf{a}_{i+d-1}$. Then, by (10), we have

$$\begin{bmatrix} Y_0(i) & \cdots & Y_{n_i}(i) \\ Y_1(i) & \cdots & Y_{n_i+1}(i) \\ \vdots & \cdots & \vdots \\ Y_{r-n_i-1}(i) & \cdots & Y_{r-1}(i) \\ \vdots & \vdots & \vdots \\ Y_0(i+d-1) & \cdots & \theta^{-n_i(d-1)} Y_{n_i}(i+d-1) \\ \theta^{-(d-1)} Y_1(i+d-1) & \cdots & \theta^{-(n_i+1)(d-1)} Y_{n_i+1}(i+d-1) \\ \vdots & \cdots & \vdots \\ \theta^{-(r-n_i-1)(d-1)} Y_{r-n_i-1}(i+d-1) & \cdots & \theta^{-(r-1)(d-1)} Y_{r-1}(i+d-1) \end{bmatrix} \begin{bmatrix} 1 \\ a_{i1} \\ \vdots \\ a_{in_i} \end{bmatrix} = \mathbf{0}. \tag{11}$$

Rewrite (11) as

$$\underbrace{\begin{bmatrix} Y_1(i) & \cdots & Y_{n_i}(i) \\ Y_2(i) & \cdots & Y_{n_i+1}(i) \\ \vdots & \cdots & \vdots \\ Y_{r-n_i}(i) & \cdots & Y_{r-1}(i) \\ \vdots & \vdots & \vdots \\ \theta^{-(d-1)} Y_1(i+d-1) & \cdots & \theta^{-n_i(d-1)} Y_{n_i}(i+d-1) \\ \theta^{-2(d-1)} Y_2(i+d-1) & \cdots & \theta^{-(n_i+1)(d-1)} Y_{n_i+1}(i+d-1) \\ \vdots & \cdots & \vdots \\ \theta^{-(r-n_i)(d-1)} Y_{r-n_i}(i+d-1) & \cdots & \theta^{-(r-1)(d-1)} Y_{r-1}(i+d-1) \end{bmatrix}}_{\mathbf{Y}_1} \underbrace{\begin{bmatrix} a_{i1} \\ a_{i2} \\ \vdots \\ a_{in_i} \end{bmatrix}}_{\mathbf{a}_i} = -\underbrace{\begin{bmatrix} Y_0(i) \\ Y_1(i) \\ \vdots \\ Y_{r-n_i-1}(i) \\ \vdots \\ Y_0(i+d-1) \\ \theta^{-(d-1)} Y_1(i+d-1) \\ \vdots \\ \theta^{-(r-n_i-1)(d-1)} Y_{r-n_i-1}(i+d-1) \end{bmatrix}}_{\mathbf{Y}_2}. \tag{12}$$

**Lemma 1**: Assume $\mathbf{I}_i = \cdots = \mathbf{I}_{i+d-1}$, $n_i = |\mathbf{I}_i| \leq N_S < r$ and the subsignals in the wideband of our interest are linearly independent. If $r \geq 2n_i$ or $d \geq n_i$, equation (10) has a unique nontrivial solution.

***Proof***: For $\mathbf{I}_i = \cdots = \mathbf{I}_{i+d-1}$, we get $\mathbf{a}_i = \cdots = \mathbf{a}_{i+d-1}$ and (10) holds. Let

$$\mathbf{Y}_1^{i+s} = \begin{bmatrix} \theta^{-s} Y_1(i+s) & \cdots & \theta^{-s(r-n_i)} Y_{r-n_i}(i+s) \\ \vdots & \ddots & \vdots \\ \theta^{-sn_i} Y_{n_i}(i+s) & \cdots & \theta^{-s(r-1)} Y_{r-1}(i+s) \end{bmatrix}_{n_i \times (r-n_i)}, s = 0, ..., d-1. \tag{13}$$

For $\theta^{-sk}\omega_{i+s,j}^{k} = e^{-j2\pi sk}e^{j2\pi(f_{i,j}+s)k} = e^{j2\pi f_{i,j}k} \triangleq \omega_{i,j}^{k}$ and by (9) we have

$$\begin{bmatrix} Y_0(i+s) \\ \theta^{-s}Y_1(i+s) \\ \vdots \\ \theta^{-s(r-1)}Y_{r-1}(i+s) \end{bmatrix} = \frac{1}{\alpha}\begin{bmatrix} 1 & \cdots & 1 \\ \theta^{-s}\omega_{i+s,0}^{1} & \cdots & \theta^{-s}\omega_{i+s,n_i-1}^{1} \\ \vdots & \ddots & \vdots \\ \theta^{-s(r-1)}\omega_{i+s,0}^{r-1} & \cdots & \theta^{-s(r-1)}\omega_{i+s,n_i-1}^{r-1} \end{bmatrix}\begin{bmatrix} X_{f_{i+s,0}} \\ X_{f_{i+s,1}} \\ \vdots \\ X_{f_{i+s,n_i-1}} \end{bmatrix} = \frac{1}{\alpha}\begin{bmatrix} 1 & \cdots & 1 \\ \omega_{i,0}^{1} & \cdots & \omega_{i,n_i-1}^{1} \\ \vdots & \ddots & \vdots \\ \omega_{i,0}^{r-1} & \cdots & \omega_{i,n_i-1}^{r-1} \end{bmatrix}\begin{bmatrix} X_{f_{i+s,0}} \\ X_{f_{i+s,1}} \\ \vdots \\ X_{f_{i+s,n_i-1}} \end{bmatrix}. \tag{14}$$

Then

$$\mathbf{Y}_1^{i+s} = \frac{1}{\alpha}\underbrace{\begin{bmatrix} \omega_{i,0}^{1} & \cdots & \omega_{i,n_i-1}^{1} \\ \omega_{i,0}^{2} & \cdots & \omega_{i,n_i-1}^{2} \\ \vdots & \ddots & \vdots \\ \omega_{i,0}^{n_i} & \cdots & \omega_{i,n_i-1}^{n_i} \end{bmatrix}}_{\boldsymbol{\omega}}\begin{bmatrix} X_{f_{i+s,0}} & \cdots & \omega_{i,0}^{r-n_i-1}X_{f_{i+s,0}} \\ X_{f_{i+s,1}} & \cdots & \omega_{i,1}^{r-n_i-1}X_{f_{i+s,1}} \\ \vdots & \ddots & \vdots \\ X_{f_{i+s,n_i-1}} & \cdots & \omega_{i,n_i-1}^{r-n_i-1}X_{f_{i+s,n_i-1}} \end{bmatrix}, \tag{15}$$

$$\mathbf{Y}_1^T = \begin{bmatrix} \mathbf{Y}_1^i, \ldots, \mathbf{Y}_1^{i+d-1} \end{bmatrix} = \boldsymbol{\omega}\underbrace{\begin{bmatrix} X_{f_{i+0,0}} & \cdots & \omega_{i,0}^{r-n_i-1}X_{f_{i+0,0}} & \cdots & X_{f_{i+d-1,0}} & \cdots & \omega_{i,0}^{r-n_i-1}X_{f_{i+d-1,0}} \\ \vdots & \cdots & \vdots & \cdots & \vdots & \cdots & \vdots \\ X_{f_{i+0,n_i-1}} & \cdots & \omega_{i,n_i-1}^{r-n_i-1}X_{f_{i+0,n_i-1}} & \cdots & X_{f_{i+d-1,n_i-1}} & \cdots & \omega_{i,n_i-1}^{r-n_i-1}X_{f_{i+d-1,n_i-1}} \end{bmatrix}}_{\mathbf{X}_\omega}_{n_i \times d(r-n_i)}. \tag{16}$$

By the property of Vandemonde matrix, $\text{rank}\{\boldsymbol{\omega}\} = n_i$. Thus,

$$\text{rank}\{\mathbf{Y}_1^T\} = \text{rank}\{\mathbf{X}_\omega\}. \tag{17}$$

Let

$$\mathbf{X}^{*} = \begin{bmatrix} X_{f_{i+0,0}} & X_{f_{i+1,0}} & \cdots & X_{f_{i+d-1,0}} \\ \vdots & \vdots & \cdots & \vdots \\ X_{f_{i+0,n_i-1}} & X_{f_{i+1,n_i-1}} & \cdots & X_{f_{i+d-1,n_i-1}} \end{bmatrix}_{n_i \times d}, \mathbf{X}^{**} = \begin{bmatrix} X_{f_{i+0,0}} & \omega_{i,0}^{1}X_{f_{i+0,0}} & \cdots & \omega_{i,0}^{r-n_i-1}X_{f_{i+s,0}} \\ \vdots & \vdots & \cdots & \vdots \\ X_{f_{i+0,n_i-1}} & \omega_{i,n_i-1}^{1}X_{f_{i+0,n_i-1}} & \cdots & \omega_{i,n_i-1}^{r-n_i-1}X_{f_{i+0,n_i-1}} \end{bmatrix}_{n_i \times (r-n_i)}.$$

Since the transmissions are linearly independent, the columns of $\mathbf{X}^{*}$ are linearly independent. It means if $d \geq n_i$, $\text{rank}\{\mathbf{X}^{*}\} = n_i$. With respect to $\mathbf{X}^{**}$, it is not hard to get that if $r \geq 2n_i$, $\text{rank}\{\mathbf{X}^{**}\} = n_i$.

For $\text{rank}\{\mathbf{X}_\omega\} \geq \max\{\text{rank}\{\mathbf{X}^{*}\}, \text{rank}\{\mathbf{X}^{**}\}\}$, if $r \geq 2n_i$ or $d \geq n_i$, we have

$$\text{rank}\{\mathbf{X}_\omega\} \geq n_i. \tag{18}$$

Note $\text{rank}\{\mathbf{X}_\omega\} \leq \min\{n_i, d(r-n_i)\}$. Thus, we have

$$\text{rank}\{\mathbf{X}_\omega\} \leq n_i. \tag{19}$$

Combining (17), (18) and (19), we have $\text{rank}\{\mathbf{Y}_1\} = \text{rank}\{\mathbf{Y}_1^T\} = n_i$, if $r \geq 2n_i$ or $d \geq n_i$. Thus, $\mathbf{Y}_1$ has full column rank and equation (12) has a unique nontrivial solution. The lemma is proved.

**Theorem 2**: Let $N_s$ be the number of the sub-bands and $r$ be the number of ADCs. Other parameters are the same as Theorem 1. If $r \geq 2N_S$ or $N_s \leq d \leq \min\{u_s - l_s, s = 1, 2, \ldots, n_D\}$, $\mathbf{I}_i, i = 0, \ldots, N/\alpha - 1$ can be uniquely determined by $n_D$ FLPs $G_i(z) = \sum_{s=0}^{u_i-l_i} a_{is}z^s, i = l_1, l_2, \ldots, l_{n_D}$, where $u_s, l_s, n_D$ are as defined in Property 1. The FLPs can be approximated by $\hat{G}_i(z) = \sum_{s=0}^{u_i-l_i} \hat{a}_{is}z^s, i = l_1, \ldots, l_{n_D}$, where $\hat{\mathbf{a}}_i = [\hat{a}_{i1}, \ldots, \hat{a}_{i,u_i-l_i}]$ is the approximation of $\mathbf{a}_i$.

*Proof*: For $n_i = |\mathbf{I}_i| \leq N_S, i = 0, \ldots, N/\alpha - 1$, then the conditions $r \geq 2N_S$ or $N_s \leq d \leq \min\{u_s - l_s, s = 1, 2, \ldots, n_D\}$ cover those in lemma 1.

In the absence of noise, we can use the similar algorithm in decoding BCH. When $d = 1, r \geq 2n_i$, we obtain the unique nontrivial solution of (12) by using Berlekamp-Massey algorithm [27]. When $d > 1$, the problem (12) is changed into a synthesis of multisequence, where we can use the algorithm presented in [28] to solve (12).

In the presence of noise, from lemma 1, if $r \geq 2N_S$ or $N_s \leq d \leq \min\{u_s - l_s, s = 1, 2, \ldots, n_D\}$, $\mathbf{Y}_1$ has full column rank. Then we can obtain the solution of (12) by least square method (LSM), where $\hat{\mathbf{a}}_i = (\mathbf{Y}_1^H \mathbf{Y}_1)^{-1} \mathbf{Y}_1^H \mathbf{Y}_2$.

The theorem is proved.

**Remark 1**: In practice, we regard that the number of subsignals is $N_S$, which is invariable, and $r = N_S + 1$ is a constant. Even through $n_i$ may be less than $N_S$, it does not affect the derivation and the conclusion of the above theorems.

## 3.3 Signal Detection Algorithm

### 3.3.1 Searching Roots of the FLPs

From section 3.1 and 3.2, we get the approximated FLPs. To obtain $\mathbf{I}_i, i = 0,...,N/\alpha-1$, we need to determine the roots of the FLPs. Let $\mathbf{Z}_i = \{z_{i,1}, z_{i,2},...,z_{i,|\mathbf{I}_i|}\}$ be the roots of the FLP corresponding to $\mathbf{H}_i$, and $\hat{\mathbf{Z}}_i = \{\hat{z}_{i,1}, \hat{z}_{i,2},...,\hat{z}_{i,|\mathbf{I}_i|}\}$ be the approximation of $\mathbf{Z}_i$. Consider the following two cases.

**Case 1**: In the absence of noise, since $\hat{G}_i(z) = G_i(z)$, $\hat{\mathbf{Z}}_i = \mathbf{Z}_i$ can be obtained by directly solve the roots of $\hat{G}_i(z)$. Here, we can use the Pan's algorithm [29] for fast finding the roots.

**Case 2**: In the presence of noise, $\hat{\mathbf{Z}}_i$ may not be accurately equal to $\mathbf{Z}_i$. We provide a simple root finding method.

Let $\mathbf{C}_i = \{e^{j2\pi f_l \tau}, f_l = i + (l-1)N/\alpha, l = 0, 1,...,\alpha-1\}$ be the candidate roots set of the FLP with respect to $\mathbf{H}_i$. Evidently, $\mathbf{Z}_i \subseteq \mathbf{C}_i$. We only need searching the roots from the $\alpha$ candidate ones. Firstly, compute the FLP evaluations of all the candidate roots and get

$$\mathbf{E}_i = \{\|\hat{G}_i(z)\| \mid z \in \mathbf{C}_i\}. \tag{20}$$

Then we obtain

$$\hat{\mathbf{Z}}_i = \{z \in \mathbf{C}_i \mid \|\hat{G}_i(z)\| \in \min_{|\mathbf{I}_i|} \mathbf{E}_i\}, \tag{21}$$

where $\min_j \mathbf{A}$ denotes the set of $j$ minimal elements of $\mathbf{A}$.

The theoretical basis of this method is: in the presence of noise, by the continuity of the complex polynomial evaluation, each root of $G_i(z)$ falls in a small neighbor area of a root of $\hat{G}_i(z)$, namely $\|\hat{G}_i(z)\| \approx 0, z \in \mathbf{Z}_i$; On the other hand, for $z \in \mathbf{C}_i \setminus \mathbf{Z}_i$, $\|\hat{G}_i(z)\| \gg 0$, which indicates that we can distinguish the true and false roots basing on their FLP evaluations. The specific decision rule is presented in the following parts.

### 3.3.2 Theoretical Preparation for Hypothesis Test

In the practical communication, noise is unavoidable. In the presence of noise, relation (12) is changed into the perturbation version

$$(\mathbf{Y}_1 + \mathbf{\Delta}_1)\hat{\mathbf{a}}_i = \mathbf{Y}_2 + \mathbf{\Delta}_2, \tag{22}$$

where $\mathbf{\Delta} = [\mathbf{\Delta}_1, -\mathbf{\Delta}_2]$ is the perturbation matrix. $\mathbf{\Delta}$ is also the FFT of the WGN added to the time-domain samples of $[\mathbf{Y}_1, -\mathbf{Y}_2]$.

We consider two cases: pure noise and noise + signal.

**Case 1**: pure noise. Firstly, give the following lemma.

**Lemma 2**: Let the real samples $x_0, x_1,..., x_{N-1}$ be zero mean white Gaussian noise (WGN) sequence with variance $\sigma_{noise}^2$. Its FFT is $X(f_k) = \sum_{n=0}^{N-1} x_n e^{-j2\pi f_k n}, k = 0,...,N-1$. Then

$$\frac{1}{\sqrt{N}} X(f) = \frac{1}{\sqrt{N}} \begin{bmatrix} X(f_1) \\ \vdots \\ X(f_{\frac{N}{2}-1}) \end{bmatrix} \sim CN(0, \sigma_{noise}^2 \mathbf{I}).$$

For complex samples, we can directly have the following corollary.

**Corollary 1**: Let $X(f_k) = \sum_{n=0}^{N-1} x_n e^{-j2\pi f_k n}, k = 0,...,N-1$ be FFT of the sequence $x_0, x_1,...,x_{N-1}$, where $x_i \sim CN(0, \sigma_{noise}^2), i = 0,...,N-1$. Then $\frac{1}{\sqrt{N}} X(f) \sim CN(0, \sigma_{noise}^2 I)$, where $X(f) = [X(f_1) \cdots X(f_{N-1})]$.

In the remainder of this paper, we redefine the FFT with $X(f_k) \triangleq \frac{1}{\sqrt{N}} \sum_{n=0}^{N-1} x_n e^{-j2\pi f_k n}$. Thus $X(f) \sim CN(0, \sigma_{noise}^2 \mathbf{I})$.

In the absence of signal, (22) is changed to

$$\mathbf{\Delta}_1 \hat{\mathbf{a}} = \mathbf{\Delta}_2. \tag{23}$$

The least square solution of (23) is

$$\hat{\mathbf{a}} = \mathbf{\Delta}_1^\dagger \mathbf{\Delta}_2 = (\mathbf{\Delta}_1^H \mathbf{\Delta}_1)^{-1} \mathbf{\Delta}_1^H \mathbf{\Delta}_2. \tag{24}$$

Let the singular value decomposition (SVD) of $\mathbf{\Delta}_1$ be $\mathbf{\Delta}_1 = \mathbf{U\Sigma V}^H$, where $\Sigma = \begin{bmatrix} \sqrt{d}\sigma_{noise} & 0 & 0 \\ 0 & \ddots & 0 \\ 0 & 0 & \sqrt{d}\sigma_{noise} \\ & 0 & \end{bmatrix}$. Then

$$\mathbf{\Delta}_1 = \sqrt{d}\sigma_{noise}\mathbf{U}\begin{bmatrix}\mathbf{I}\\0\end{bmatrix}\mathbf{V}^H \quad \text{and}$$

$$\hat{\mathbf{a}} = (\mathbf{\Delta}_1^H\mathbf{\Delta}_1)^{-1}\mathbf{\Delta}_1^H\mathbf{\Delta}_2 = (d\sigma_{noise}^2)^{-1}\mathbf{\Delta}_1^H\mathbf{\Delta}_2. \tag{25}$$

By corollary 1, the components of $\mathbf{\Delta}$ are independent, and $\mathbf{\Delta}_{ij} \sim CN(0,\sigma_{noise}^2)$. Hence,

$$E\{\hat{\mathbf{a}}_i\} = (d\sigma_{noise}^2)^{-1} E\{\mathbf{\Delta}_1^H\} E\{\mathbf{\Delta}_2\}$$
$$\text{var}\{\hat{\mathbf{a}}_i\} = (d\sigma_{noise}^2)^{-2}\text{var}\{\mathbf{\Delta}_1^H\}\text{var}\{\mathbf{\Delta}_2\} = (d\sigma_{noise}^2)^{-2}(d\sigma_{noise}^4\mathbf{I}) = d^{-1}\mathbf{I} \tag{26}$$

Here, the components of $\hat{\mathbf{a}}_i = [\hat{a}_{i1},\hat{a}_{i2},...,\hat{a}_{in_i}]$ obey the complex Gaussian distribution, namely $\hat{a}_{ik} \sim CN(0,d^{-1}), k = 1,...,n_i$. The FLP evaluations (FLPEs) of the candidate roots are

$$\hat{G}_i(z) = 1 + [z, z^2,..., z^{n_i}]\hat{\mathbf{a}}_i, z \in \mathbf{C}_i.$$

Since $\|z\| = 1, z \in \mathbf{C}_i$, the FLPE $\hat{G}'_i(z) = \hat{a}_{i1}z + \hat{a}_{i2}z^2 + ... + \hat{a}_{in_i}z^{n_i}$ can be regarded as a special kind DFT of $(a_{i1},...,a_{in_i})$. By corollary 1, we have $G'_i(z) \sim CN(0, n_i d^{-1})$. It is not hard to get that the real and imaginary part of $G'_i(z)$ are independent. Then we have $E(\hat{G}_i(z)) = E(1+zG'_i(z)) = 1, \text{var}(\hat{G}_i(z)) = n_i d^{-1}$, namely,

$$\hat{G}_i(z) \sim CN(1, n_i d^{-1}), z \in \mathbf{C}_i. \tag{27}$$

Relation (27) show that the FLPE of the candidate roots are unrelated to the noise level and only determined by $d$.

**Theorem 3**: In the absence of signal, the FLPE $\hat{G}_i(z) \sim CN(1, n_i d^{-1}), z \in \mathbf{C}_i$. Namely, $\text{Re}\{\hat{G}_i(z)\} \sim N(1, n_i d^{-1}/2)$ and $\text{Im}\{\hat{G}_i(z)\} \sim N(0, n_i d^{-1}/2)$. $\|\hat{G}_i(z)\| \sim Rice(1, \sqrt{n_i d^{-1}/2}), z \in \mathbf{C}_i$, where $Rice(\cdot)$ denotes the Rice distribution.

Fig. 2 gives the histgrams of the FLPEs from 1000 times tests for pure noises of different levels. The figures show that the variances of the evaluations in the two simulations are equal, which coincides with the theoretical results. Since $\frac{1}{n_i d^{-1}/2}$ is relatively large, $\|\hat{G}_i(z)\|$ has an approximated Gaussian distribution, which is demonstrated by Fig. 2. In the following part of this paper, we regard $\|\hat{G}_i(z)\| \sim N(1, n_i d^{-1}/2)$.

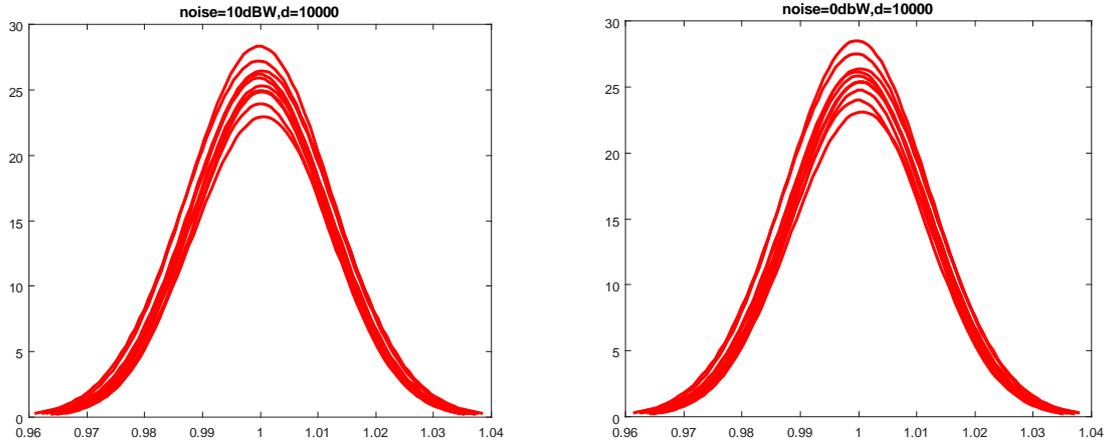

Fig. 2 Histgrams of FLPEs (absolute value) of the candidate roots. Test runs for 1000 times at aliased frequency 3MHz for $n_i = 3$. The variances in the two cases are respectively 1.4961e-04 and 1.4893e-04 where the theoretical value is 1.5000e-04.

**Case 2:** Noise + signal.

Relation of each row of (12) can be characterized by $\mathbf{a}_i = [a_{i1}, a_{i2},..., a_{in_i}]$, which is the coefficients of the frequency locator polynomial. The continuity of the communication signal in the frequency domain guarantees that these linear relations determine the identical FLP. The scale of the relations is determined by the bandwidths of the subsignals. The FLPEs of the candidate roots has the following properties.

Let $\mathbf{a} = [a_1, a_2,..., a_{n_i}]$ be the actual coefficient of a certain FLP. $\hat{\mathbf{a}}$ is the least square approximation of $\mathbf{a}$. Assume

$$\hat{\mathbf{a}} = \mathbf{a} + \tilde{\mathbf{a}}. \tag{28}$$

By (22) we have

$$(\mathbf{Y}_1 + \mathbf{\Delta}_1)\hat{\mathbf{a}} \approx \mathbf{Y}_2 + \mathbf{\Delta}_2, \tag{29}$$

where $\mathbf{\Delta} = [\mathbf{\Delta}_1, -\mathbf{\Delta}_2]$ is the perturbation matrix. Obviously,

$$\mathbf{Y}_1\mathbf{a} = \mathbf{Y}_2. \tag{30}$$

Combining (28) and (29), we have

$$(\mathbf{Y}_1 + \mathbf{\Delta}_1)(\mathbf{a} + \tilde{\mathbf{a}}) \approx \mathbf{Y}_2 + \mathbf{\Delta}_2. \tag{31}$$

By LSM and (29), we get

$$\hat{\mathbf{a}} = \left( \left( \mathbf{Y}_1 + \boldsymbol{\Delta}_1 \right)^H \left( \mathbf{Y}_1 + \boldsymbol{\Delta}_1 \right) \right)^{-1} \left( \mathbf{Y}_1 + \boldsymbol{\Delta}_1 \right)^H \left( \mathbf{Y}_2 + \boldsymbol{\Delta}_2 \right). \tag{32}$$

For $z \in \mathbf{Z}_i$, let $\mathbf{z} = (z, z^2, ..., z^{n_i-1})$. Then

$$\begin{aligned}
\mathbf{z}\tilde{\mathbf{a}} &= \mathbf{z}\left( \left( \mathbf{Y}_1 + \boldsymbol{\Delta}_1 \right)^H \left( \mathbf{Y}_1 + \boldsymbol{\Delta}_1 \right) \right)^{-1} \left( \mathbf{Y}_1 + \boldsymbol{\Delta}_1 \right)^H \left( \mathbf{Y}_2 + \boldsymbol{\Delta}_2 \right) - \mathbf{z}\mathbf{a} \\
&= \mathbf{z}\left( \left( \mathbf{Y}_1 + \boldsymbol{\Delta}_1 \right)^H \left( \mathbf{Y}_1 + \boldsymbol{\Delta}_1 \right) \right)^{-1} \left( \mathbf{Y}_1 + \boldsymbol{\Delta}_1 \right)^H \left( \mathbf{Y}_1 + \boldsymbol{\Delta}_1 \right) + 1 \\
&= \mathbf{z}\left( \left( \mathbf{Y}_1 + \boldsymbol{\Delta}_1 \right)^H \left( \mathbf{Y}_1 + \boldsymbol{\Delta}_1 \right) \right)^{-1} \left( \mathbf{Y}_1^H \mathbf{Y}_1 \right) \mathbf{a} + \mathbf{z}\left( \left( \mathbf{Y}_1 + \boldsymbol{\Delta}_1 \right)^H \left( \mathbf{Y}_1 + \boldsymbol{\Delta}_1 \right) \right)^{-1} \boldsymbol{\Delta}_1^H \mathbf{Y}_2 \\
&\quad + \mathbf{z}\left( \left( \mathbf{Y}_1 + \boldsymbol{\Delta}_1 \right)^H \left( \mathbf{Y}_1 + \boldsymbol{\Delta}_1 \right) \right)^{-1} \left( \mathbf{Y}_1 + \boldsymbol{\Delta}_1 \right)^H \boldsymbol{\Delta}_2 + 1
\end{aligned} \tag{33}$$

The distribution of $\tilde{\mathbf{a}}$ is relatively complicated. Instead, we give an approximation. Assume that $d$ is sufficiently large. Using SVD, we have

$$\mathbf{A} \triangleq \left( \mathbf{Y}_1 + \boldsymbol{\Delta}_1 \right)^H \left( \mathbf{Y}_1 + \boldsymbol{\Delta}_1 \right) = U \begin{pmatrix} \sigma_1^2 + \sigma_{noi}^2 & 0 & 0 \\ 0 & \ddots & 0 \\ 0 & 0 & \sigma_{n_i}^2 + \sigma_{noi}^2 \end{pmatrix} U^H, \tag{34}$$

where $\sigma_1^2, ..., \sigma_{n_i}^2$ are the singular values contributed by the signal whereas $\sigma_{noi}^2 \approx d\sigma^2$ is dominated by the noise.

i. In the low SNRs, $\sigma_{n_i}^2 \leq \cdots \leq \sigma_1^2 \ll \sigma_{noi}^2$. Here, we have an approximation that $\sigma_{n_i}^2 \approx \cdots \approx \sigma_1^2 \approx \sigma_{sig}^2$. Then

$$\mathbf{A} \approx \begin{pmatrix} \sigma_{sig}^2 + \sigma_{noi}^2 & 0 & 0 \\ 0 & \ddots & 0 \\ 0 & 0 & \sigma_{sig}^2 + \sigma_{noi}^2 \end{pmatrix}, \tag{35}$$

and

$$\left( \left( \mathbf{Y}_1 + \boldsymbol{\Delta}_1 \right)^H \left( \mathbf{Y}_1 + \boldsymbol{\Delta}_1 \right) \right)^{-1} \mathbf{Y}_1^H \mathbf{Y}_1 \approx \begin{pmatrix} \frac{\sigma_{sig}^2}{\sigma_{sig}^2 + \sigma_{noi}^2} & 0 & 0 \\ 0 & \ddots & 0 \\ 0 & 0 & \frac{\sigma_{sig}^2}{\sigma_{sig}^2 + \sigma_{noi}^2} \end{pmatrix}. \tag{36}$$

Combine (33) and (36), then

$$\begin{aligned}
\mathbf{z}\tilde{\mathbf{a}} &\approx \mathbf{z}\left( \left( \mathbf{Y}_1 + \boldsymbol{\Delta}_1 \right)^H \left( \mathbf{Y}_1 + \boldsymbol{\Delta}_1 \right) \right)^{-1} \left( \mathbf{Y}_1^H \mathbf{Y}_1 \right) \mathbf{a} + \mathbf{z}\left( \left( \mathbf{Y}_1 + \boldsymbol{\Delta}_1 \right)^H \left( \mathbf{Y}_1 + \boldsymbol{\Delta}_1 \right) \right)^{-1} \boldsymbol{\Delta}_1^H \mathbf{Y}_2 + \mathbf{z}\left( \left( \mathbf{Y}_1 + \boldsymbol{\Delta}_1 \right)^H \left( \mathbf{Y}_1 + \boldsymbol{\Delta}_1 \right) \right)^{-1} \mathbf{Y}_1^H \boldsymbol{\Delta}_2 + 1 \\
&\approx \frac{\sigma_{sig}^2}{\sigma_{sig}^2 + \sigma_{noi}^2} \mathbf{z}\mathbf{a} + \mathbf{z}\left( \left( \mathbf{Y}_1 + \boldsymbol{\Delta}_1 \right)^H \left( \mathbf{Y}_1 + \boldsymbol{\Delta}_1 \right) \right)^{-1} \boldsymbol{\Delta}_1^H \mathbf{Y}_2 + \mathbf{z}\left( \left( \mathbf{Y}_1 + \boldsymbol{\Delta}_1 \right)^H \left( \mathbf{Y}_1 + \boldsymbol{\Delta}_1 \right) \right)^{-1} \mathbf{Y}_1^H \boldsymbol{\Delta}_2 + 1 \\
&\approx \frac{-\sigma_{sig}^2}{\sigma_{sig}^2 + \sigma_{noi}^2} + \frac{1}{\sigma_{sig}^2 + \sigma_{noi}^2} \mathbf{z}\boldsymbol{\Delta}_1^H \mathbf{Y}_2 + \frac{1}{\sigma_{sig}^2 + \sigma_{noi}^2} \mathbf{z}\mathbf{Y}_1^H \boldsymbol{\Delta}_2 + 1
\end{aligned}$$

Hence,

$$\begin{aligned}
E(\mathbf{z}\tilde{\mathbf{a}}) &\approx \frac{\sigma_{noi}^2}{\sigma_{sig}^2 + \sigma_{noi}^2} = \frac{1}{1 + SNR} \\
\text{var}(\mathbf{z}\tilde{\mathbf{a}}) &\approx \|\mathbf{Y}_2\|_F^2 \frac{1}{(\sigma_{sig}^2 + \sigma_{noi}^2)^2} n_i \frac{\sigma_{noi}^2}{d} + \|\mathbf{Y}_2\|_F^2 \frac{1}{(\sigma_{sig}^2 + \sigma_{noi}^2)^2} n_i \frac{\sigma_{noi}^2}{d} \\
&\approx \frac{2\|\mathbf{Y}_2\|_F^2 n_i \sigma_{noi}^2}{d(\sigma_{sig}^2 + \sigma_{noi}^2)^2} \approx \frac{2\sigma_S^2 n_i \sigma_{noi}^2}{d(\sigma_{sig}^2 + \sigma_{noi}^2)^2} = \frac{2 n_i \sigma_{sig}^2 \sigma_{noi}^2}{d(\sigma_{sig}^2 + \sigma_{noi}^2)^2} \\
&= n_i d^{-1} \frac{2 SNR}{(1 + SNR)^2}
\end{aligned} \tag{37}$$

Where $\|\cdot\|_F$ denotes the Frobenius norm.

**Theorem 4**: In the low SNRs, namely $\sigma_{n_i}^2 \leq \cdots \leq \sigma_1^2 \ll \sigma_{noi}^2$, where $\sigma_{n_i}^2, ..., \sigma_1^2, \sigma_{noi}^2, \sigma_{sig}^2$ is as defined in (36). For $z \in \mathbf{Z}_i$, $\hat{G}(z) = 1 + \mathbf{z}\hat{\mathbf{a}} = 1 + \mathbf{z}\tilde{\mathbf{a}} + \mathbf{z}\mathbf{a} = \mathbf{z}\tilde{\mathbf{a}}$ and approximately

$$\hat{G}(z) \sim CN\left( \frac{1}{1 + SNR}, \frac{2n_i d^{-1} SNR}{(1 + SNR)^2} \right), \tag{38}$$

Accordingly, $\text{Re}\{\hat{G}(z)\} \sim N\left( \frac{1}{1 + SNR}, \frac{n_i d^{-1} SNR}{(1 + SNR)^2} \right)$, $\text{Im}\{\hat{G}(z)\} \sim N\left( 0, \frac{n_i d^{-1} SNR}{(1 + SNR)^2} \right)$ and $\|\hat{G}_i(z)\| \sim Rice(\frac{1}{1 + SNR}, \sqrt{\frac{n_i d^{-1} SNR}{(1 + SNR)^2}})$.

Here, we also regard that $\|\hat{G}_i(z)\| \sim N(\frac{1}{1+SNR}, \frac{n_i d^{-1} SNR}{(1+SNR)^2}), z \in \mathbf{Z}_i$. For $z \in \mathbf{C}_i \setminus \mathbf{Z}_i$, the distribution is as (27).

ii. In the high SNRs, $\sigma_1^2 \geq \cdots \geq \sigma_{n_i}^2 \gg \sigma_{noi}^2 \approx 0$. In this case, $\tilde{\mathbf{a}} \approx \mathbf{0}, \hat{\mathbf{a}} \approx \mathbf{a}$. Then, $\hat{G}(z) \approx 0$ for $z \in \mathbf{Z}_i$ and $\hat{G}(z) \gg 0$ for $z \in \mathbf{C}_i - \mathbf{Z}_i$.

### 3.3.3 Multiband Signal Detector

From Section 3.3.2, we get that in the high SNRs, the FLPEs (absolute value) of the true roots are nearly zero, whereas the evaluations of the false roots are far larger than zero. It is not complicated to decide the absence or presence of the signals basing on the FLPEs of the candidate roots. Hence, we mainly pay attention to the case of low SNRs.

Theorem 3 and 4 respectively present the distribution of the FLP evaluations of the candidate roots in the case of pure noise and that of the simultaneous presence of noise and signals. Thus $\|\hat{G}(z)\|$ can be selected as the detection measurement. Consider that we have no priori knowledge about the number of nonzero frequencies in each bucket in the practical case. In the detector, we set up the value to its maximum $N_S$. Hence, the detection measurements are $\|\hat{G}_i(z^*)\|, z^* \in \hat{\mathbf{Z}}_i$, where $\hat{\mathbf{Z}}_i$ is the set of candidate roots which have the minimum FLPE. Then

$$\begin{cases} \mathbf{H}1: \mathrm{Re}\{\hat{G}_i(z^*)\} < \xi, & \text{frequency is occupied} \\ \mathbf{H}0: \mathrm{Re}\{\hat{G}_i(z^*)\} > \xi, & \text{frequency is unoccupied} \end{cases}, \quad (51)$$

Where $\xi = \sqrt{n_i/d}(1+Q^{-1}(1-P_F))$ is the detection threshold, $P_F$ is the false alarm ratio and the corresponding detection probability is $P_d \approx 1 - Q(\frac{\xi}{\sqrt{n_i d^{-1} SNR/(1+SNR)^2}} - \frac{1}{1+SNR})$.

The signal detecting process includes three steps:

**Step 1**: choose the $N_S$ roots of minimal FLP evaluations as the preliminary actual roots;

**Step 2**: use (51) to decide whether the roots obtained in the step 1 indicate the nonzero frequencies or not;

**Step 3**: combine the decisions in Step 1 and 2, then get $\mathbf{I}_i, i = 0, \ldots, N/\alpha - 1$. Basing on $\mathbf{I}_i$, estimate the number of sub-band signals and their parameters.

### 3.3.4 Estimating the Bandwidths and the Carrier Frequencies

By the detector presented in Section 3.3.3, we have already obtained the frequency location sets $\mathbf{I}_i, i = 0, \ldots, N/\alpha - 1$ corresponding to $\mathbf{H}_i$. Then we can get $\hat{\mathbb{S}} = \bigcup_{i=0}^{N/\alpha-1} \mathbf{I}_i = \bigcup_{k=0}^{\hat{n}_S - 1} [\hat{l}_k, \hat{u}_k)$, which is the estimation of signal support $\mathbb{S} = \bigcup_{k=0}^{n_S - 1} [l_k, u_k)$, where $[l_k, u_k)$ and $[\hat{l}_k, \hat{u}_k)$ respectively denote the frequency locations of the $k-th$ subsignal $s_k(t)$ and its estimation. Directly get the bandwidths $b_k = \hat{u}_k - \hat{l}_k$ and carrier frequencies $\tilde{f}_k^c = \frac{\hat{u}_k + \hat{l}_k}{2}$ from $\hat{\mathbb{S}}$.

Note the carrier frequencies are the values obtained after frequency conversion by RF-fronts. The actual values should add the parts of converted frequencies.

### 3.3.5 Detection Algorithm

The whole signal detection algorithm is shown in Alg. 1. Here we only consider the cases that $n_i = N_S = r - 1, i = 0, \ldots, N/\alpha - 1$ and the other cases are considered in the next part. In Alg. 1, the mapping $e(x) = \exp(j2\pi\tau x/N)$ is used.

---

**Algorithm 1: Multi-band signal detection**

**Input**: $y_1(n), y_2(n), \ldots, y_r(n), N, \alpha, n \in [0, N-1]$, $i = 0, \mathbf{I}_{-1} \triangleq \text{null}, \hat{\mathbb{S}} = \text{null}$

**Output**: $(b_k, f_k^c), k = 1, 2, \ldots, N_S$

compute the FFT of the subsampled signals and obtain
$Y_1(m), Y_2(m), \ldots, Y_r(m), m \in [0, N/\alpha - 1]$.

**while** $i < N/\alpha$ **do**

    $\hat{\mathbf{a}}_i \leftarrow \mathbf{Y}1^\dagger \mathbf{Y}2$

    $\hat{G}_i(z) = 1 + [z^1, \ldots, z^{r-1}] \hat{\mathbf{a}}_i$

    $\mathbf{E}_i \leftarrow \{\|\hat{G}_i(z)\|^2, z \in \mathbf{C}_i\}$

    $\hat{\mathbf{Z}}_i \leftarrow \arg\min_{z \in \mathbf{C}_i}^{r-1} \mathbf{E}_i$

    **if** $\|\hat{G}_i(z)\| < \xi, z \in \hat{\mathbf{Z}}_i$ **then**

        $\hat{\mathbf{I}}_i \leftarrow \hat{\mathbf{I}}_i \cup \{e^{-1}(z)\}$

    **end if**

```
Î_i ← Î_i ∪ Î_{i-1}
Ŝ ← Ŝ ∪ Î_i
i ← i+1
end while
compute b_k and f_k^c with Ŝ for k = 1, 2, ..., n_S
return (b_k, f_k^c), k = 1, 2, ..., n_S
```

### 3.3.6 Solving the Boundary Problem

The previous sections only consider processing the support recovery in every district defined in Property 1, while the recovery at the borders between the adjoining districts are not considered. We call it the boundary problem.

By Property 1 we know that the buckets in a certain districts determine an identical FLP. However, in the adjoining parts, the bins correspond to at least two different FLPs. Namely, different FLPs may be corresponding to the relation of each row in (12). We note that relation (12) is derived from (9). In fact, the boundary problem originates from that some components of $\mathbf{X}_i$ in (9) may be zero. This does not affect the followed derivation. It only results that the detector reports the locations of all nonzero frequencies aliased in the boundary districts. For example, the number of the reported nonzero frequencies at the border between $\mathbf{H}_i$ and $\mathbf{H}_{i+1}$ are the sum of those of both buckets.

Fig. 3 shows the decision results in the boundary districts. There are three signals S1, S2 and S3 marked in different colors aliasing at frequency from f0 to f4. $[0, f4+d]$ is divided into 15 districts. The decision results are placed at the center right above each districts, e.g. S1+S2 means the nonzero frequencies aliased in this district are contributed by signal S1 and S2. By the results, we can estimate the carrier frequencies and bandwidths. For example, the frequency where the decisions including S2 ranges at $[f1-d, f4]$, then we get S2 occupies the buckets $\mathbf{H}_{f1 \sim f4}$. The carrier of S2 is $(f4+f1)/2$ and the bandwidth is $f4-(f1-d)-d = f4-f1$.

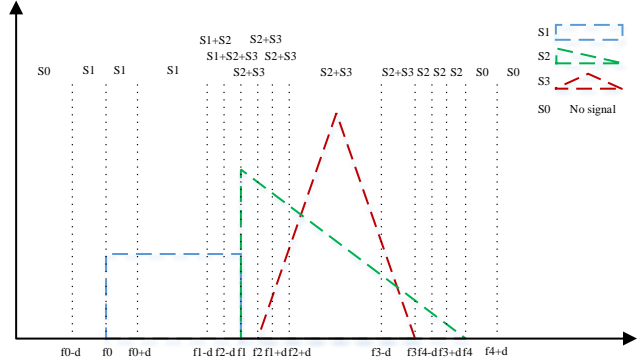

Fig. 3 Decision results for the boundary districts.

## 4 Comparison

The advantages of the proposed sparse multiband signal detection algorithm includes the following three aspects: using sub-Nyquist sampling method, which can alleviate the burden on ADC; low complexity of both samples and computation; high detection probability in the low SNRs, which is then demonstrated by the simulation results.

Table 1 presents the comparison between different detection methods. The matched filter has the highest accuracy but requires full priori knowledge. In contrast with MF, energy detection need no priori knowledge and has the lowest computing complexity, however, its performance deteriorates in the low SNRs. In this section, we mainly compare our algorithm with cyclostationary detection.

Table1 Comparison between Detection Methods

| Method | Priori Knowledge | Accuracy | Complexity | | Robustness to Noise |
|---|---|---|---|---|---|
| | | | Samples | Computing | |
| Matched filter | full | highest | Nyquist rate | high | yes |
| Energy Detection | no | low | Nyquist rate | low | no |
| Cyclostationary Detection | no | high | $8/5 f_{lan}$ (sparse)<br>$4/5 f_{nyq}$ (no constraints) | high | yes |
| FLP Based Algorithm | no | high | $(N_S+1)/N_S f_{lan}$ | low | yes |

### 4.1 Sampling Method

In our scheme, we use multi-coset sampling method. The time shifts of the sampling channels satisfy $\tau_s = s, s = 0, ..., r-1$, such that the subsampled signals are restricted by the linear relation (12). A common RF-front is used in MC, whereas MWC has more RF-fronts. In the sense of hardware complexity, MC has superiority. On the other hand, MC is limited by the performance of a single RF-front. The analog input bandwidth of a RF-front limits the maximal bandwidth that our algorithm can sense. For broader use in the future, we need to design a suited RF-fronts circuit. We note that our algorithm owns the potential to use the sampling scheme of MWC to get a similar spectrum aliasing as MC and arithmetic distributed time shifts.

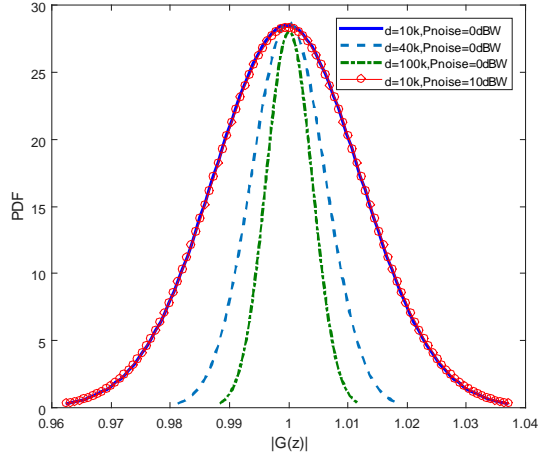

Fig. 4 Histfits of the PDF of $\text{Re}\{G(z)\}$ (pure noise) for different $d$ and noise power. The variances of the four PDFs are respectively 0.3033e-04, 0.0801e-04, 0.3111e-05, 0.3058e-04.

**4.2 Complexity**

The minimal sampling rate of the algorithm proposed in [19] is $8/5 f_{lan}$ (sparse case) or $4/5 f_{nyq}$ (no constraints), where $f_{lan} = N_S B_{\max}$ is the Landau rate. The total sampling rate of our method is $(N_S+1)B_{\max}$, namely $f_{lan}(N_S+1)/N_S$. When the number of the subbands $N_S > 1$ and the wideband spectrum is sparse, the sample complexity of our algorithm is less than that of [19].

The computing burden for cyclostationary detection [19] gathers in the process using CTF and OMP. The running time of OMP is $O(K^2 N \ln N)$ [31]. The CTF block need to find a sparsest solution matrix which is an NP-hard problem [32], which can also be solved by using OMP algorithm. When the spectrum support is changeless, the CTF block can run for only once. However, CTF will run again once the support changes. With respect to our algorithm, the running time is dominated by the resolution of (12). The cost of solving (12) with LSM is $O((d+N_S)N_S^2)$. This operation handles for $\frac{N}{\alpha d}$ times. Considering $d \gg N_S$, then the total running time is $O\left(\frac{(d+N_S)N_S^2 N}{\alpha d}\right) = O\left(\frac{d N_S^2 N}{\alpha d}\right) = O\left(N_S \frac{N_S N}{\alpha}\right) = O(K)$, where $K = \frac{N_S N}{\alpha} = N_S B_{\max}$ is the spectral sparsity.

**4.3 Robustness to Noise**

The cyclostationary detection has strong robustness to noise since the cyclic spectrum of noise at nonzero cyclic frequencies are zero. If the detected signal is cyclostationary, then the algorithm in [19] can robustly detect the signal. In contrast, our algorithm does not restrict the signal in cyclic property. It only requires that the minimal bandwidth of the subband signals is greater than the number of the transmissions. The robustness of the proposed algorithm is demonstrated in the simulation results.

## 5 Simulation Results

We now demonstrate the multiband signal detection and investigate the performance of the proposed detector via simulations. We also compare our approach to energy detection. Throughout the simulations, we suppose the maximal bandwidth of the subsignals satisfies $B_{\max} \leq N/\alpha$ and the transmission number $N_S \leq r-1$. The analog signal is sampled by multi-coset. The relative intervals of the sampling channels are $\tau_s = s, s = 0,...,r-1$. Only four ADCs (analog input: 100MHz, sampling rates: 10MS/s) are used. Thus, $N = f_{nyq} = 100\text{M}, \alpha=10, r=4$. Since the original signal is real, the spectrum is conjugate symmetric. Only half of the spectrum is considered and the original signals are generated such that the number of the nonzero frequencies aliased in each point on $[0,5\text{M}]$Hz does not exceed 3. Thus $N_S = 3$ and $B_{\max} = 5\text{MHz}$.

We consider the additive Gaussian white noise, the SNR is defined as the ratio of the power of the original signal and the added noise, namely

$$\text{SNR} = \frac{\sum_{i=0}^{N_S-1} \|s_i(t)\|^2}{\|n(t)\|^2}.$$

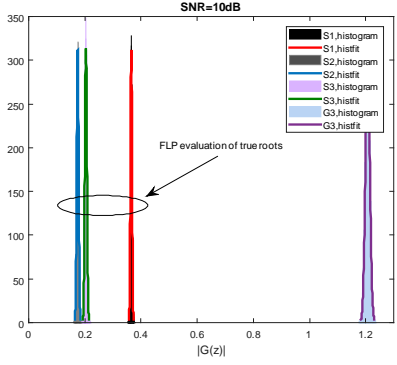 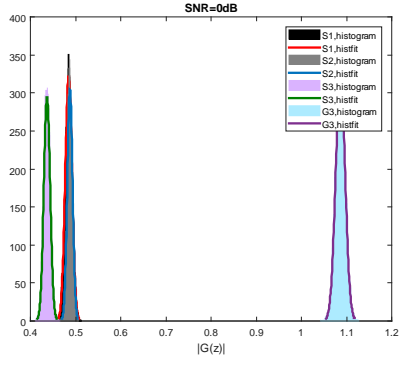 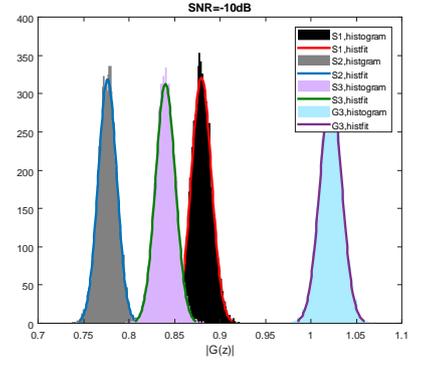

Fig a. SNR=10dB　　　　　　　　　　　　Fig b. SNR=0dB　　　　　　　　　　　　Fig c. SNR=-10dB

Fig. 5 Histograms of the PDF of $\text{Re}\{G(z)\}$ for different SNRs. Only the PDFs of four FLP evaluations with minimal absolute values are shown, where S1, S2, S3 denote the evaluations of the roots corresponding to the nonzero frequencies dominated by the occupied bands and G3 represents the minimal evaluation (absolute value) of the false roots. The target aliased frequency is 3.0 MHz and the original signal are composed of three sub-signals, of which the carriers are 23 MHz, 33 MHz, 43 MHz and the bandwidths are all 3 MHz. The simulation runs for 10000 times.

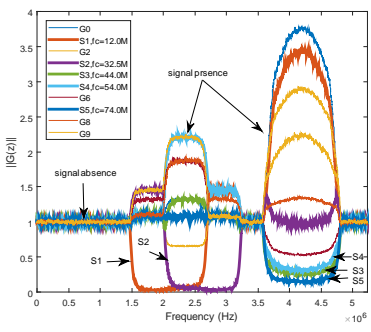 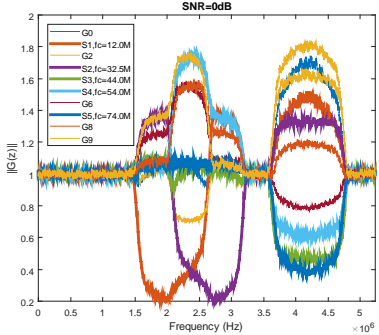 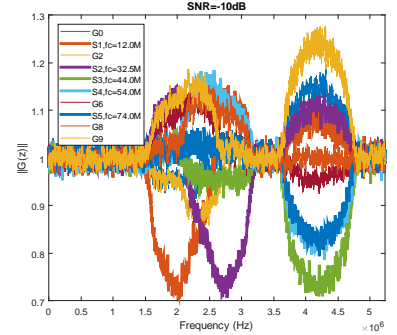

Fig 6. FLP evaluation for all the candidate roots in different SNRs. In the simulations, $d = 10000$. $(G0,...,G9) = (G(w^0),...,G(w^9))$, where $w^s = e^{j2\pi f_s/N}$, $f_s = i + sN/\alpha$ for $i \in [0, 5M)$. $f_c$ denotes the carrier frequency and S1 ~ S5 denote the FLP evaluations of the true roots indicating the locations of the occupied bands. The occupied bands are [11.5M, 12.5], [32M, 33M], [43.5M, 44.5M], [53.5M, 54.5M] and [73.5M, 74.5M].

## 5.1 Numerical results of the FLP Evaluation

To demonstrate the exactness of the theoretical results, we firstly investigate the PDF of the FLP evaluation for practical data by histogram. Fig. 4 depicts the histfits of $\text{Re}\{G(z)\}$ for different pure noise power and different $d$, where $z$ is one of the candidate roots. Here $z = e^{j2\pi f/N}, f = 22.5M$ is selected. The simulation runs for 1000 times. Only the curves fitting the histograms are shown. The results show that both mean and variance of $\text{Re}\{G(z)\}$ accord well with the theoretical values, which are 1 and $d^{-1}n_i/2 = 1.5000e-04$ respectively, where $n_i = 3$ for all the simulations. We also note that the variance of FLPE is only related to $d$ and $n_i$. Thus the first and fourth curves in Fig. 4 coincide.

Fig. 5 presents the PDFs of $\text{Re}\{G(z)\}$ for four of the candidate roots, where three of them are corresponding to the nonzero frequencies. The mean and variance of $\text{Re}\{G(z)\}, z \in \mathbf{Z}_i$ conform to the theoretical results given in Theorem 4. With the SNR decreasing, $\text{Re}\{G(z)\}, z \in \mathbf{C}_i$ all converge to 1. From Fig. 5 we note that even when SNR=-10dB, there exists a gap between the evaluations of true roots and the other candidate ones.

The roots indicating the locations of occupied bands are closely neighboring on the unit for the simulations in Fig. 5. In this case, $\min \|G(z)\| = 8\sin\frac{\pi}{10}\sin\frac{\pi}{10}\sin\frac{2\pi}{10} = 1.1756, z \in \mathbf{C}_i \setminus \mathbf{Z}_i$ in the absence of noise. Accordingly, the global minimal $\|G(z)\|$, $z \in \mathbf{C}_i \setminus \mathbf{Z}_i$ for $n_i = 1, 2, 3$ are respectively 0.6180, 0.3820 and 0.4490. We call the cases that the minimum $\|G(z)\|$ exists for $z \in \mathbf{C}_i \setminus \mathbf{Z}_i$, the 'worst' cases, where the false alarm ratios are larger than those of other cases. The worst cases have been discussed in [21].

Fig 6 gives the FLPE $\|G(z)\|$ for all candidate roots throughout the 5M buckets. The original signal is composed of five[2] transmissions with carriers 12.0 MHz, 32.5 MHz, 44.0 MHz, 54 MHz, 74 MHz and bandwidths of each 1 MHz. Obviously, the simulations simultaneously suffer the three worst cases, where $\min\|G(z)\| = 0.6180$, $\min\|G(z)\| = 0.3820, z \in \mathbf{C}_i \setminus \mathbf{Z}_i$ and $\min\|G(z)\| = 0.4490$ respectively for the buckets in the districts $[1.5M, 2M) \cup [2.5M, 3M)$, $[2M, 2.5M)$ and $[3.5M, 4.5M]$.

---

[2] The number of subbands is five which is more than $N_S = 3$. However, for each bucket, $n_i \leq N_S = 3$ still holds. Therefore, our algorithm still works in this case.

Since the practical algorithm regards $n_i \equiv 3$, only the worst case $\min\|G(z)\| = 0.6180$ ('G2') and $\min\|G(z)\| = 0.4490$ ('G6') can be seen from Fig 6.

We note that the roots corresponding to 'G2' and 'G6' in Fig 6 may produce false alarm. In fact, this kind of false alarms can be avoided by reverse proving. Here, we take 'G2' as an example. We firstly suppose that 'G2' corresponds to the true roots, then 'G2' along with 'S1' and 'S2' should have the PDFs as S1, S2 and S3 displayed in Fig 5. If the assumption stands, then 'G2' should have the same PDF as S2 in Fig 5. Namely, 'G2' should have the FLP evaluation (absolute value) less than or nearly equal to the 'S1' and 'S2'. However, this does not coincide with the results presented in Fig 6. Thus, the supposition does not hold and 'G2' can be judged as the false roots.

To demonstrate the feasibility of this method more clearly, we give the average evaluation weights for the three possible frequency collisions. The results are depicted in Fig. 7. The SNR is set to -10 dB. From Fig. 7, the overall FLP evaluation weights of all the candidate roots for different numbers of aliased frequencies differs from each other. We have the following conclusions.

i. In the case that three frequencies hashed to a buckets fig. 7a), the evaluation of the fourth root is lower than the third and the fifth ones, while larger when two nonzero frequencies collide (fig. 7c).

ii. In both the cases, the three evaluations of the third, fourth and the fifth roots are distinctly lower than those of other candidate roots.

iii. In the case that only one nonzero frequency is hashed in the bucket (fig 7b), the evaluation of only one candidate root is clearly lower than those of the others. The candidate root of lowest FLP value indicates the frequency location of the signal.

Through the method provided above, we can get rid of the 'G2'-like false alarms, the decision threshold is easy to choose according to the required false alarm probability. We mark this method as 'new method'.

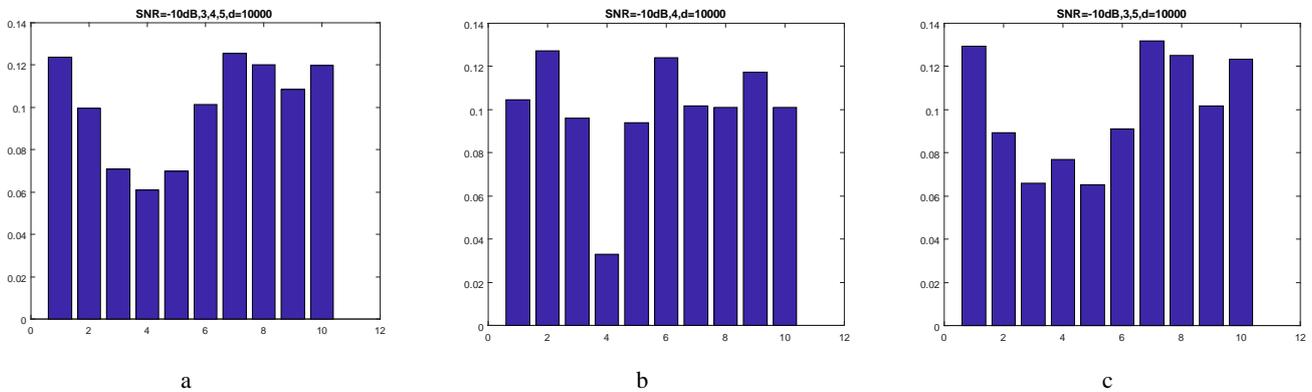

Fig. 7 Evaluation weights. The results are the average of 1000 tests.

**5.2 Multiband Signal Detection**

Next, we perform signal detection on the obtained FLP evaluations. We compare the performance of the proposed algorithm and energy detection respectively performed on the polynomial evaluations and the reconstructed power spectrum. In the experiments, we consider the multiband signal composed of three QPSK modulated sub-band signals, of which the carrier frequencies are 32.5MHz, 42.5MHz and 72.5MHz and the bandwidths are all 3MHz. We address a blind scenario where the carrier frequencies of the signals are unknown to the algorithms. The detection is defined as the occupied frequencies are correctly detected; the false alarm is declared if a detection is claimed at the unoccupied frequency. The total number of the occupied and unoccupied frequencies are respectively 9M and 41M.

The receiver operating characteristic (ROC) curve is shown in Fig. 7. Fig. 7.a shows the ROC for different SNRs and $d$. The detection probability decreases with the SNR decreasing. However, even when SNR=-10dB, the detection ratio is still relatively high. We also investigate the influence of $d$ to the performance of the detector. We set $d=1K$ and $d=10K$ for SNR=-10dB. The simulation results are depicted in Fig. 7.a. It is clear that larger $d$ results in higher detection ratio. Numerous experimental results suggest that $d=10K$ is sufficient for the detector even in very low SNRs.

Fig. 7.b presents the ROC curve for different kinds of detector. In the figure, LS-FLP and TLS-FLP respectively denote the proposed methods based on LS and TLS (total least squares), which are used in solving the problem (12). From the curves we get that the LS based and TLS based algorithm have the approximate performance. We can also see that our algorithm outperforms ED in both SNR=0dB and SNR=-10dB. Here, we note that the detection probability of ED presented in this paper may be higher than that reported in other researches. This is because ED is performed on the signal support recovered by using the FLP based method, which is the key technology of this paper and can robustly recovered the support.

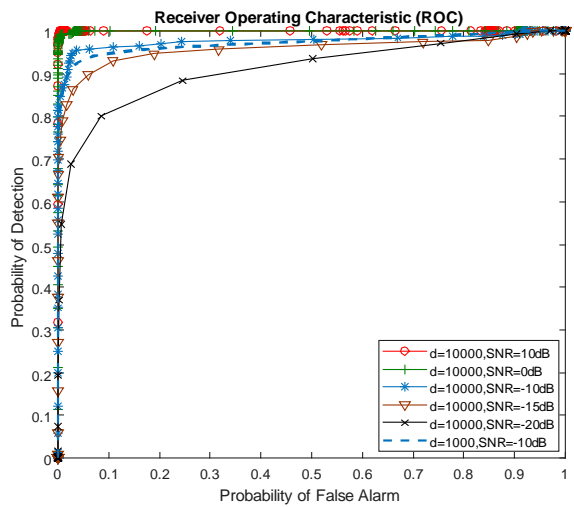 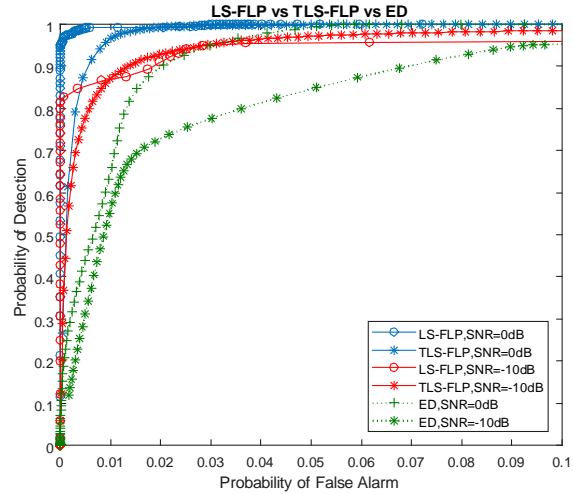

a. ROC for the proposed detection algorithm        b. ROC for LS-based, TLS-based and ED algorithm

Fig. 7 Receiver operating characteristic (ROC) curve

### 5.3 Carrier Frequencies and Bandwidths Estimation

In this simulation, we address the problem of estimating the carrier frequencies and bandwidths of the signals of interest. The target multiband signal includes $N_s = 3$ sub-signals, whose carriers are $f_1^c = 21.5\text{MHz}$, $f_2^c = 64.0\text{MHz}$, $f_3^c = 64.0\text{MHz}$, and bandwidths are $B_1 = B_2 = B_3 = 2\text{MHz}$. SNR is set to -5dB. The experimental results are: $\hat{N}_s = 3$, $\hat{f}_1 = 21.47\text{MHz}$, $\hat{f}_2 = 42.13\text{MHz}$, $\hat{f}_3 = 64.01\text{MHz}$, $\hat{B}_1 = 2.01\text{MHz}$, $\hat{B}_2 = 1.98\text{MHz}$ and $\hat{B}_3 = 1.99\text{MHz}$.

## 6 Conclusion

In this paper we focus on the blind detection of sparse multiband signal in the sub-Nyquist regime. We propose a simple and practical detection algorithm based on frequency locator polynomial whose roots indicate the locations of the occupied bands. Our algorithm can be regarded as a new kind of detection algorithm beyond matched filter, energy detection and cyclostationary detection. Theoretical analysis shows that the proposed algorithm has lower complexity of both sample and computation. Experimental results demonstrate the better performance compared to energy detection. Our method performs well in the low SNRs and. Moreover, sub-Nyquist sampling scheme is used thus to alleviate the burden of the ADCs.

As mentioned in the paper, the maximal sensed bandwidth of our method is limited by the analog input bandwidth of a single RF-front. In the future work, we will focus on redesigning the RF-fronts similar to or based on MWC, which is a potential technic popularly used in compressed sensing.